\documentclass[aps,prb,twocolumn,showpacs,superscriptaddress]{revtex4-1}

\usepackage{amsfonts}
\usepackage{amsmath}
\usepackage{amssymb}
\usepackage{graphicx}
\usepackage{bm}

\begin{document}


\title{Evolution of normal and superconducting properties of single crystals of Na$_{1-\delta}$FeAs upon interaction with environment}

\author{M.~A.~Tanatar}
\email[Corresponding author: ]{tanatar@ameslab.gov}
\affiliation{Ames Laboratory, Ames, Iowa 50011, USA}

\author{N.~Spyrison}

\affiliation{Department of Physics and Astronomy, Iowa State University, Ames, Iowa 50011, USA }

\author{Kyuil Cho}

\affiliation{Ames Laboratory, Ames, Iowa 50011, USA}
\affiliation{Department of Physics and Astronomy, Iowa State University, Ames, Iowa 50011, USA }

\author{E.~C.~Blomberg}

\affiliation{Ames Laboratory, Ames, Iowa 50011, USA}
\affiliation{Department of Physics and Astronomy, Iowa State University, Ames, Iowa 50011, USA }

\author{ Guotai Tan }

\affiliation{ Department of Physics and Astronomy, The University of Tennessee, Knoxville, Tennessee 37996-1200, USA}
\affiliation{ College of Nuclear Science and Technology, Beijing Normal University, Beijing 100875, China }

\author{ Jiaqiang Yan }

\affiliation{ Department of Physics and Astronomy, The University of Tennessee, Knoxville, Tennessee 37996-1200, USA}

\author{ Pengcheng Dai }

\affiliation{ Department of Physics and Astronomy, The University of Tennessee, Knoxville, Tennessee 37996-1200, USA}

\author{ Chenglin Zhang }

\affiliation{ Department of Physics and Astronomy, The University of Tennessee, Knoxville, Tennessee 37996-1200, USA}

\author{R.~Prozorov}
\email[Corresponding author: ]{prozorov@ameslab.gov}
\affiliation{Ames Laboratory, Ames, Iowa 50011, USA}
\affiliation{Department of Physics and Astronomy, Iowa State University, Ames, Iowa 50011, USA }

\date{\today}

\begin{abstract}
Iron-arsenide superconductor Na$_{1-\delta}$FeAs is highly reactive with the environment. Due to the high mobility of Na ions, this reaction affects the entire bulk of the crystals and leads an to effective stoichiometry change. Here we use this effect to study the doping evolution of normal and superconducting properties of \emph{the same} single crystals. Controlled reaction with air increases the superconducting transition temperature, $T_c$, from the initial value of 12~K to 27~K as probed by transport and magnetic measurements. Similar effects are observed in samples reacted with Apiezon N-grease, which slows down the reaction rate and results in more homogeneous samples. In both cases the temperature dependent resistivity, $\rho_a(T)$, shows a dramatic change with exposure time. In freshly prepared samples, $\rho_a(T)$ reveals clear features at the tetragonal-to-orthorhombic ($T_s$ $\approx$ 60~K) and antiferromagnetic ($T_m$=45~K) transitions and superconductivity with onset $T_{c,ons}$=16~K and offset $T_{c,off}$=12~K. The exposed samples show $T-$linear variation of $\rho_a(T)$ above $T_c,ons$=30~K ($T_{c,off}$=26~K), suggesting bulk character of the observed doping evolution and implying the existence of a quantum critical point at the optimal doping. The resistivity for different doping levels is affected below $\sim$200~K suggesting the existence of a characteristic energy scale that terminates the $T-$linear regime, which could be identified with a pseudogap.
\end{abstract}





\maketitle



\section{Introduction}

\begin{figure}[tbh]%
\includegraphics[width=8cm]{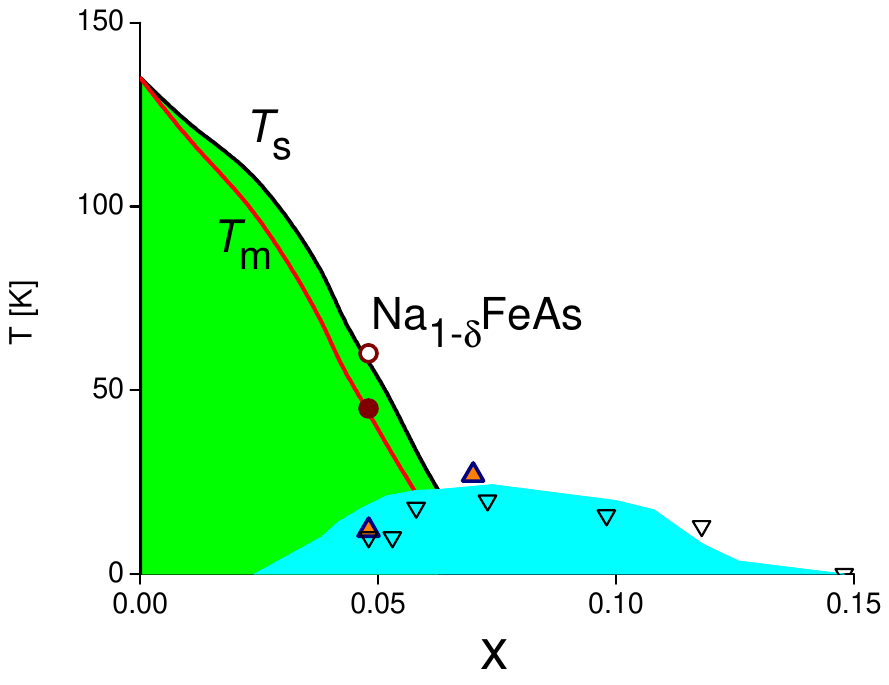}%
\caption{(Color online) Generic temperature-composition phase diagram of electron-doped iron arsenide superconductors, exemplified by Ba(Fe$_{1-x}$Co$_x$)$_2$As$_2$.\cite{Pratt} Lines of structural, $T_s$, and magnetic, $T_m$, transitions are split with Co-doping and superconductivity has maximum $T_c$ at $x_{opt} \approx $0.07, close to the composition where $T_s(x)$ and $T_m{x}$ extrapolate to zero. NaFeAs in its parent state is representative of an underdoped part of the phase diagram and the temperatures of its split structural and magnetic transitions (circles) correspond to Co-doped Ba122 with $x$=0.048. With this $x$-axis shift, actual Co-doping phase diagram for Na Fe$_{1-x}$Co$_x$As (down-triangles, data from Ref.~[\onlinecite{NaFeAsPRL104}]) matches closely the phase diagram of Ba(Fe$_{1-x}$Co$_x$)$_2$As$_2$. Up-triangles show $T_c$ and tentative $x$ position of the samples in which doping is achieved by Na deintercalation (this study), working as an electron dopant, moves the sample towards optimal doping conditions. }%
\label{phaseD}%
\end{figure}

In the generic phase diagram of FeAs-based materials, domain of superconductivity has maximum $T_c$ close to a quantum critical point of the magnetic order, see Fig.~\ref{phaseD}, suggestive of a magnetically mediated mechanism of superconductivity.\cite{Monthoux,MazinNature} Quantum fluctuations lead to a characteristic $T-$linear temperature dependence of electrical resistivity, $\rho_a(T)$, and its evolution towards a Fermi-liquid $T^2$ behavior with doping (see Ref.~[\onlinecite{Louis}] for a review). Both these features are observed in iron pnictides, most clearly in the electron doped Ba(Fe$_{1-x}$Co$_x$)$_2$As$_2$ (BaCo122 in the following) \cite{NDL} and in isoelectron doped BaFe$_2$(As$_{1-x}$P$_x$)$_2$.\cite{MatsudaQCP} In hole-doped (Ba$_{1-x}$K$_x$)Fe$_2$As$_2$ (BaK122, in the following)\cite{Rotter} resistivity also shows a limited range of close to $T-$linear variation above $T_c$, however, terminated at high-temperatures by a tendency to saturation. \cite{Wencrystals,Zverev} Similar slope changes are found in both in-plane and inter-plane resistivity of self-doped LiFeAs \cite{LiFeAsAPLKorea,TanatarLiFeAs}, and is a dominant feature in the inter-plane resistivity, $\rho _c(T)$, of transition metal doped BaTM122\cite{pseudogap,pseudogap2} where, by comparison with NMR data\cite{NMRPG}, it was associated with the formation of a pseudogap. Similar association was suggested for the explanation of both in- and inter-plane resistivity in BaK122.\cite{BaKresistivity}

Na$_{1-\delta}$FeAs (Na111 in the following) is isostructural to LiFeAs (Li111 in the following); however, in its stoichiometric form, $\delta = 0$, it is representative of an underdoped part of the superconducting phase diagram, revealing coexisting superconducting ($T_c$=12~K), magnetic ($T_m$=45~K), and orthorhombic ($T_s$=55~K) phases. \cite{NaFeAsPRL102,NaFeAsPRL104,NaFeAsNMR,NaFeAsneutrons} In line with this conclusion, superconductivity in Na111 can be enhanced with pressure \cite{pressure-NaFeAs} and Co-doping. \cite{NaFeAsPRL104} The phenomenology of split structural and magnetic transitions and coexisting superconductivity is reminiscent of BaCo122.\cite{NiNiCo,Pratt} In contrast with Li111, Na111 can be prepared with Na deficiency, allowing for some control of the Na stoichiometry.\cite{Chu,Todorov} Due to the high mobility of Na atoms at room temperature, Na can be removed from the sample by controlled oxidative deintercalation reaction with the environment.\cite{Todorov} Here we use this effect to study the evolution of the magnetic susceptibility and the temperature-dependent in-plane resistivity, $\rho_a(T)$, in \emph{the same} single crystals of NaFeAs as a function of Na content. Our main finding is the observation of $T-$linear resistivity at the optimal doping achieved by the environmental deintercalation, suggesting the universality of a quantum critical scenario in iron-pnictide superconductors. We find that the temperature range of this $T-$linear dependence is bound from the high-temperature side by the pseudogap feature, quite similar to BaK122. The pseudogap affects $\rho_a(T)$ even in the parent composition.

\section{Experimental}

Single crystals of NaFeAs were grown by sealing a mixture of Na,
Fe and As together in Ta tubes and cooking at 950 C , followed by 5 C/hour
cool-down to 900 C.\cite{crystalgrowthPRL105}
Samples were stored and transported in sealed containers filled with an inert gas. Preparation of samples for dipper tunnel diode resonator (TDR) and resistivity measurements was done in air as quickly as possible (typically about 5 minutes). The leftovers of unreacted sodium cause intense reaction on the surface of some samples. However, cleaved internal parts of single crystals do not show any visible reaction and turns out to be relatively stable. After sample preparation, samples were measured and stored in a dry, inert environment.

\begin{figure}[tbh]%
\includegraphics[width=8cm]{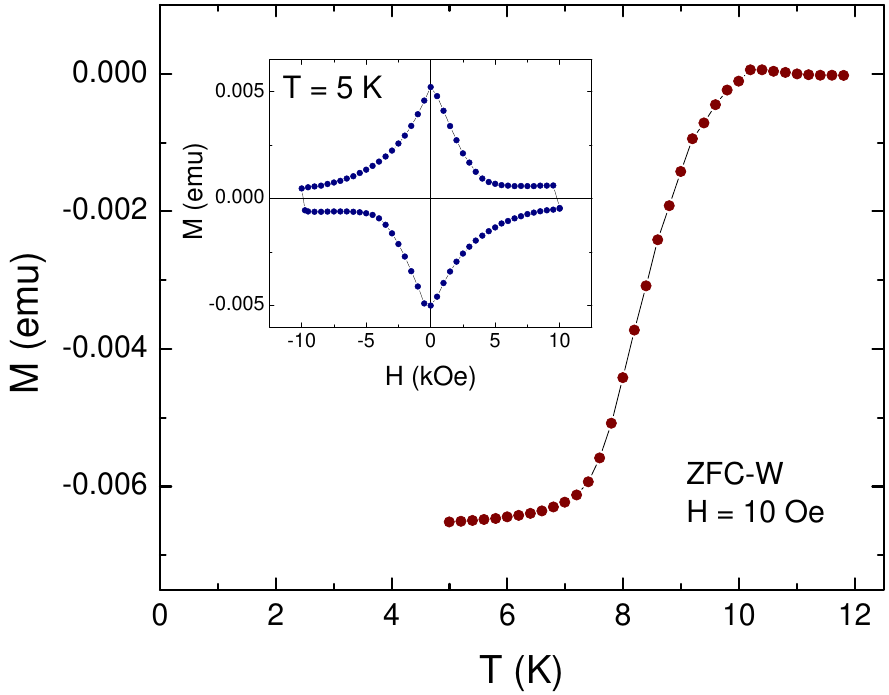}%
\caption{(Color online) DC magnetization of fresh cleaved sample measured upon warming after cooling in zero magnetic field and applying a 10 Oe field (ZFC-W). Inset: magnetic hysteresis loop measured at 5 K in the same sample.}%
\label{mag}%
\end{figure}

Samples for in-plane resistivity measurements had typical dimensions of (1- 2)$\times$0.5$\times$(0.02-0.1) mm$^3$. All sample dimensions were measured with an accuracy of about 10\%. Contacts for four-probe resistivity measurements were made by soldering 50 $\mu$m silver wires with ultra-pure Sn solder, as described in Ref.~[\onlinecite{SUST}]. This technique produced contact resistance typically in the 10 $\mu \Omega$ range. Resistivity measurements were performed in {\it Quantum Design} PPMS.

After initial resistivity measurements, samples were left mounted on a PPMS puck in air and measurements were repeated after several controlled exposures to air. Some samples, after initial preparation, were covered with Apiezon N-grease, a technique which was used to preserve Li111 in our previous studies.\cite{HyunsooLiFeAs,TanatarLiFeAs} In the case of Na111 we found that Apiezon N-grease does not prevent the sample from losing Na, but slows it down giving a better control of sample property variation, especially in combination with the ultrasonic treatment.

DC magnetic measurements were performed with a \emph{Quantum Design} MPMS on a freshly cleaved sample. AC magnetic characterization was performed with a tunnel-diode resonator (TDR)\cite{Vandegrift,Prozorov2000}, - a self-oscillating tank circuit powered by a properly biased tunnel diode. Samples were mounted on a sapphire rod and inserted in the inductor coil. The measured frequency shift is proportional to differential magnetic susceptibility of the sample.\cite{Prozorov2000} For quick mounting and measurement protocols, we used a simplified version of the TDR susceptometer (a ``dipper''), which is inserted directly into the transport $^4$He dewar. The trade-off is reduced stability and higher temperature-dependent background as compared to our high-stability $^3$He and dilution refrigerator versions of the TDR susceptometer. Nevertheless, the ``dipper'' is perfectly suitable to study magnetic signature of the superconducting transition, especially when a quick turn-around is required.

\section{Results}

Figure \ref{mag} shows the temperature - dependent DC magnetization of a freshly cleaved sample measured upon warming after cooling in zero magnetic field and applying a 10 Oe field (ZFC-W). The magnetic hysteresis loop measured at 5 K in the same sample is shown in the inset. Both results are quite characteristic of bulk superconductivity showing robust screening and significant pinning. Moreover, as we show below by direct polarized-light imaging and resistivity measurements, this bulk superconductivity coexists with structural and magnetic long-range order.

\begin{figure}[tbh]%
\includegraphics[width=8cm]{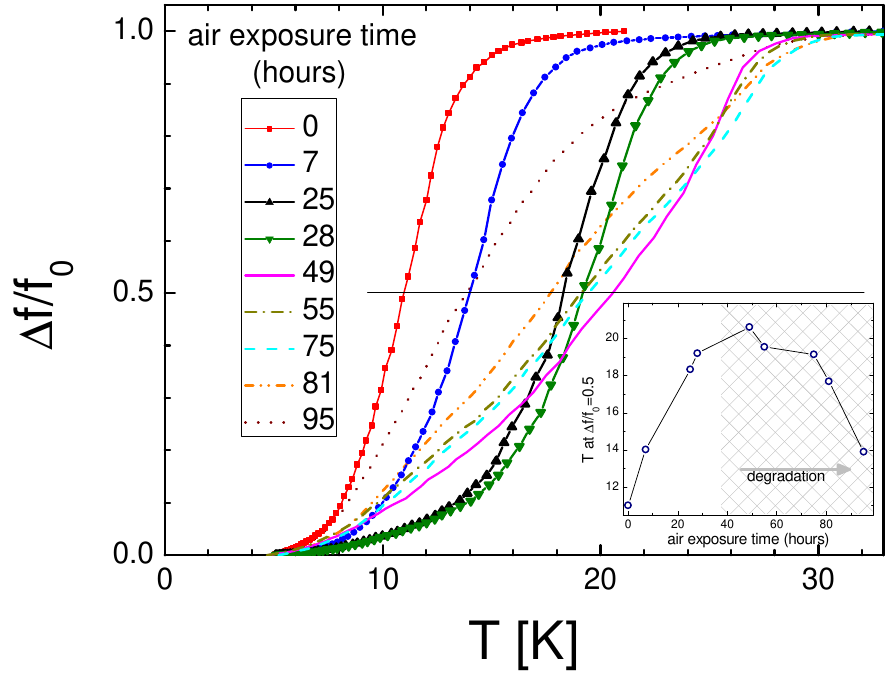}%
\caption{(Color online) Temperature dependence of the normalized frequency shift in the TDR experiment ( $f_0$ is the frequency shift at $T_c$.), for sample of NaFeAs exposed to air for the indicated number of hours. Oxidative deintercalation of Na increases the onset temperature of the superconducting transition from 13~K, for fresh samples, to 24~K for samples exposed for 28~hours in air (curves with symbols), only slightly broadening the transition. On further exposure, the onset $T_c$ rises to almost 28~K, however, additional features and significant broadening in the temperature dependence of the frequency shift show that the sample becomes inhomogeneous. Inset shows the temperature at half of the transition where $\Delta f/f_0=0.5$. The shaded area represents proliferation of degradation in the exposed samples.}%
\label{TDRair}%
\end{figure}

Figure~\ref{TDRair} shows the temperature dependence of the normalized frequency shift (proportional to magnetic susceptibility) in single crystal sample of Na111 as a function of exposure time to air. Here $f_0$ is the frequency shift at $T_c$. After initial preparation, the sample shows a quite sharp superconducting transition with the onset at $T_c \sim$13~K. Air exposure for up to 28 hours shifts the transition to $T_c \sim$24~K, almost parallel to the initial curve and does not broaden the transition. This observation suggests that the effect is truely bulk in nature, since close to $T_c$, where the London penetration depth diverges, the penetration of rf field into the sample is determined by the smaller of two characteristic length scales, - skin depth in the normal state and characteristic sample dimension, both of the order of 100 $\mu$m in this case. Upon further exposure to air, the onset temperature of the superconducting transition continues to rise until reaching the maximum value of $T_c \sim$28~K, however, the transition broadens, and additional features start to appear at lower temperatures, suggesting that the sample becomes inhomogeneous and starts to decompose into NaFe$_2$As$_2$.\cite{Todorov} The experiment with air exposure was repeated on a total of five samples and consistently produced the same result.

\begin{figure}[tbh]%
\includegraphics[width=8cm]{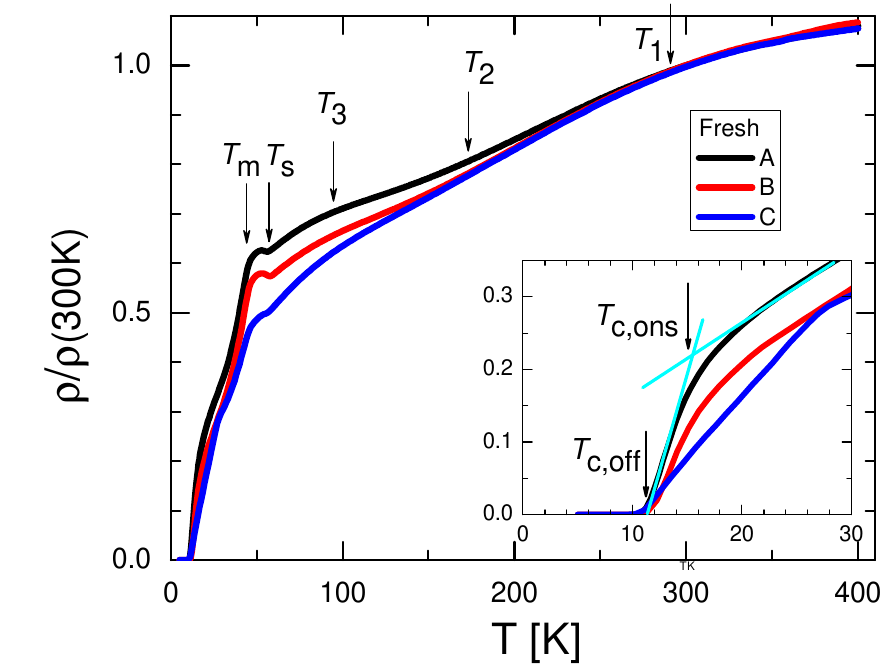}%
\caption{(Color online) Temperature dependence of the normalized in-plane resistivity, $\rho_a / \rho _a(300K)$ for three samples of NaFeAs in a fresh state after initial sample handling and contact making. Arrows show features in the temperature dependence due to magnetic and structural transitions, and additional broad slope change features at temperatures $T_3 \approx$100~K, $T_2\approx$180~K and $T_1\approx$290~K. The inset zooms onto the superconducting transition and shows the definitions of onset and offset criteria for the  superconducting transition temperature. }%
\label{samples}%
\end{figure}

Temperature-dependent resistivity of three ``fresh'' crystals of Na111 is shown in Fig.~\ref{samples} using a normalized resistivity scale, $\rho/\rho (300K)$. The resistivity value at room temperature, $\rho _a(300K)$, shows notable scatter between 200 to 400 $\mu \Omega $cm,  presumably due to the effect of hidden cracks on the effective sample geometric factor.\cite{NiNiCo,anisotropy} These values of $\rho _a(300K)$ are comparable within the error bars to the value of $\sim$300 $\mu \Omega $cm, typical for parent \cite{Alloul,anisotropy2} and hole-doped BaK122 compounds \cite{BaKresistivity,BaKresistivityWen}, as well as for Li111 \cite{HyunsooLiFeAs}, and is notably lower than the previously reported value of above 4500 $\mu \Omega $cm.\cite{NaFeAsPRL102} The temperature dependence of the resistivity of the three fresh samples A, B and C is very close, with some difference potentially coming from the difference in time of air exposure during sample preparation and contact making. The air exposure can also be responsible for smoother features in the temperature-dependent resistivity in this study compared to the previous study. \cite{NaFeAsPRL102} On cooling, $\rho _a (T)$ of fresh samples reveals three broad resistivity slope changes at $T_1 \sim$ 290~K, $T_2 \sim$180~K and $T_3 \sim$ 90~K, a small $\rho_a(T)$ upturn on cooling below $T_s \sim$56~K, a sharp downturn below $T_m \sim$45~K and a superconducting transition with onset at $T_{c,ons}\sim$~14~K (sample A) and offset at $T_{c,off}\sim$12~K. Both onset and offset temperatures were determined as crossing points of linear $\rho_a(T)$ extrapolations above, at, and below the transition as shown for $T_{c,ons}$ in inset in Fig.~\ref{samples}. The onset temperature of resistive transition in sample C shifts to 26~K, showing that it is indeed more exposed to initial degradation during sample preparation.

\begin{figure}[tbh]%
\includegraphics[width=8cm]{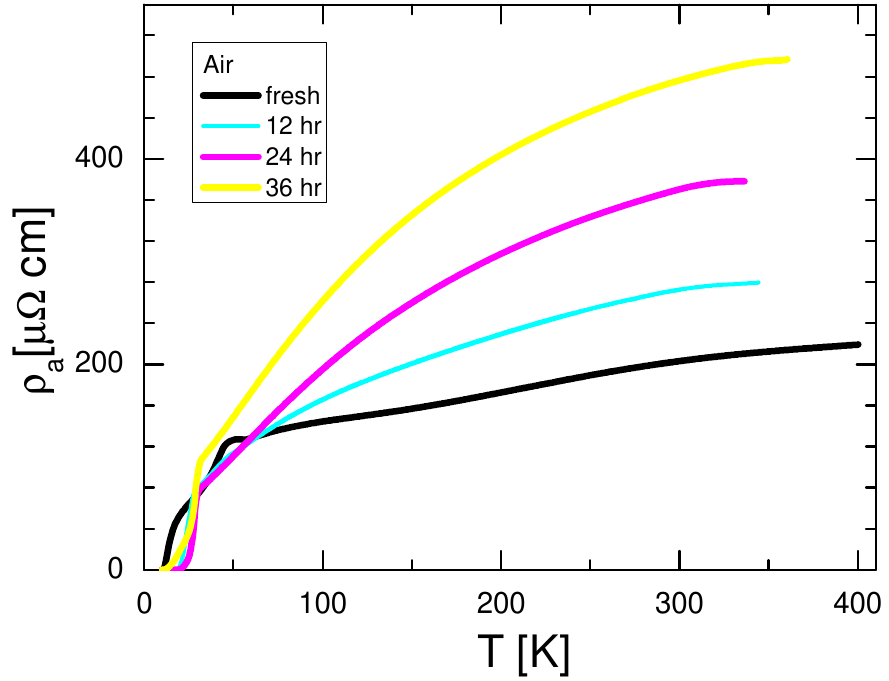}\\*
\includegraphics[width=8cm]{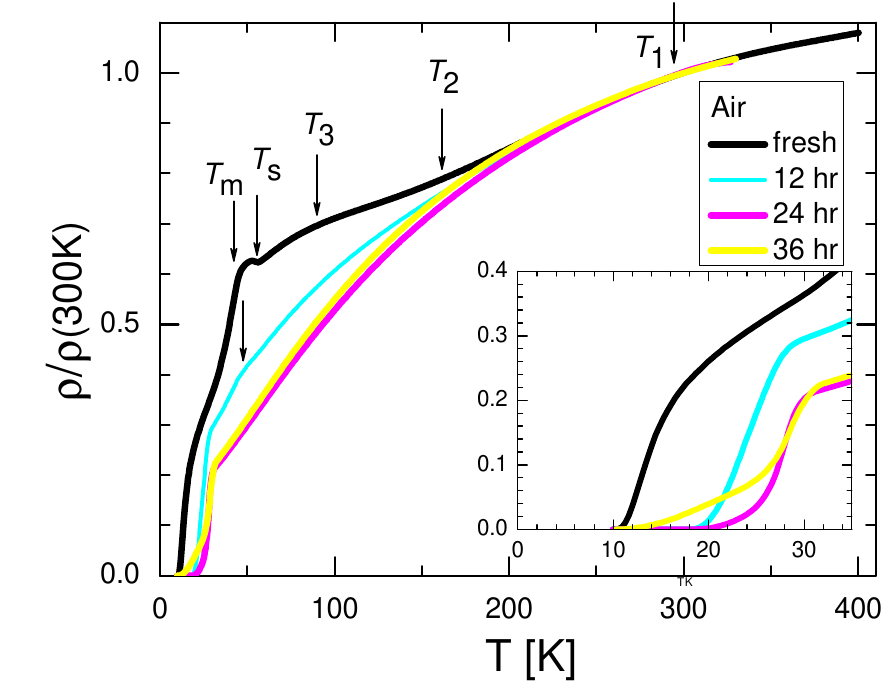}
\caption{(Color online) (Top panel) The evolution of the temperature-dependent resistivity of the sample of Na$_{1-\delta}$FeAs during oxidative deintercalation of Na upon exposure to air. Resistivity of the sample at room temperature, $\rho_a (300K)$, monotonically increases due to macroscopic sample degradation and formation of cracks. Bottom panel shows the same data, plotted using a normalized resistivity scale, $\rho / \rho (300K)$. The normalization procedure removes variation of the effective geometric factor and essentially reveals doping - independent resistivity close to room temperature and a strong variation below 200~K. The inset in the bottom panel zooms onto the superconducting transition range. Arrows show positions of the special features in $\rho_a(T)$ in the parent and 12 hour exposed samples, showing suppression of the structural/magnetic transition temperature with doping and evolution of the crossover features.
}%
\label{resistivityair}%
\end{figure}

Fig.~\ref{resistivityair} shows the evolution of the temperature-dependent resistivity, $\rho _a(T)$, in samples of Na111 during exposure to air with total exposure time up to 36~hours, during which sample properties change homogeneously in the TDR experiments, see Fig.~\ref{TDRair}. The top panel shows the evolution of sample resistivity, while the bottom panel shows the same data plotted using a normalized resistivity scale, $\rho/\rho (300K)$, and a zoom of the superconducting transition area. It is clear that the environmental reaction changes the bulk properties of the sample, grossly affecting its normal state $\rho _a(T)$. On the other hand, air exposure monotonically increases $\rho _a(300K)$ , suggesting development of macroscopic defects (most likely cracks) during Na deintercalation. Plotting the data on the normalized scale removes the variation of the geometric factor due to cracks and reveals very peculiar transformation of the temperature-dependent resistivity. First, all four curves coincide above approximately 200~K, consistent with the idea that the variation of $\rho _a(300K)$ value comes predominantly from the variation of the geometric factor and that the variation of doping level plays only a minor role. Second, below 200~K the $\rho _a(T)$ curve systematically decreases with air exposure. Indeed, even unscaled resistivity values decrease causing curves crossing in Fig.~\ref{resistivityair}b. With the exposure time of 12 hours, the small resistive increase at $T_s$ becomes indistinguishable, while the rapid downturn below $T_m$ shifts to lower temperatures as expected for the increased doping level, and $T_{c,ons}$ rises to 30~K while $T_{c,off}$ rises to 20~K. For a 24 hour exposure time, the $\rho_a(T)$ curve shows the sharpest superconducting transition with the onset at 30~K, crosses zero at 26~K, and a tail at the end of the transition with actual zero reached at 20~K. With further exposure increase to 36 hours, the tail belowthe  transition develops further, in broad accordance of the characteristic time scale with sample degradation in TDR measurements, Fig.~\ref{TDRair}.

Together with suppression of the structural transition and an increase of the superconducting transition, the crossover features in resistivity at $T_3$ become completely indistinguishable in a 12 hour curve, while the resistivity variation through $T_2$ changes from a slight slope decrease on cooling in fresh samples to the onset of a rapid resistivity decrease in exposed samples. Interestingly, $\rho_a (T)$ becomes $T$-linear above $T_c$ in samples with a 24 hour exposure, while the high-temperature end of the $T$-linear range is close to $T_2$.

\begin{figure}[tbh]%
\includegraphics[width=8cm]{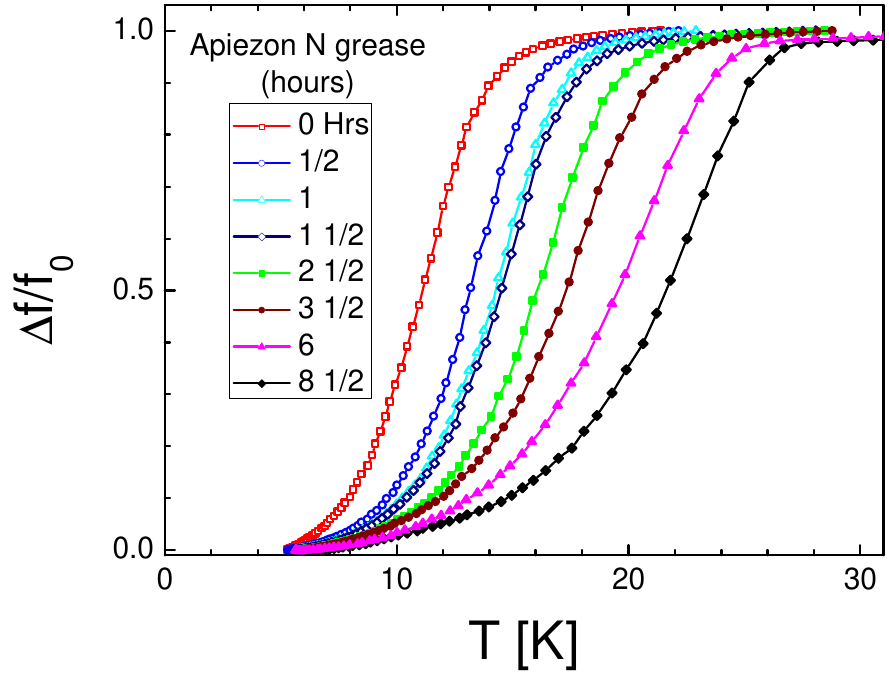}%
\caption{(Color online) Temperature dependence of the normalized frequency shift in the TDR experiment for a sample of Na$_{1-\delta}$FeAs stored in Apiezon N-grease and treated in an ultrasonic cleaner between successive runs. Oxidative deintercalation of Na increases the onset temperature of the superconducting transition from 13~K for fresh samples to 26~K with a slight transition broadening, however, without the appearance of additional features observed in air-exposed samples, Fig.~\ref{TDRair}. }%
\label{TDRgrease}%
\end{figure}

In Fig.~\ref{TDRgrease} we show the evolution of the TDR magnetic susceptibility in samples stored in Apiezon N-grease and subjected to ultrasonic treatment to facilitate Na diffusion between dipper runs. While grease protects samples from air, it does not stop Na deintercalation, which now leads to a much higher quality of the superconducting transition without additional features and only minor broadening.

\begin{figure}[tbh]%
\includegraphics[width=8cm]{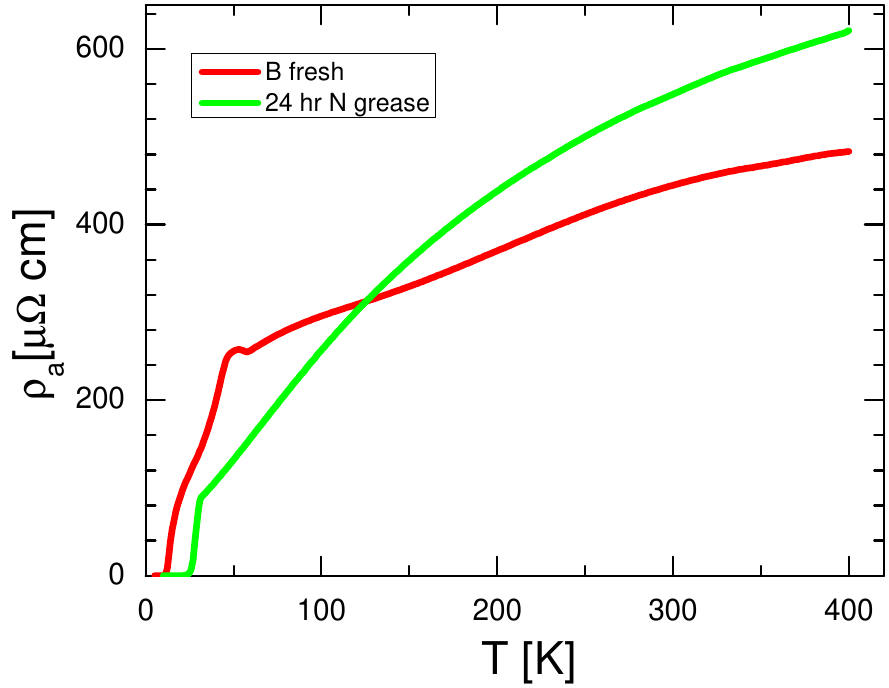}\\*
\includegraphics[width=8cm]{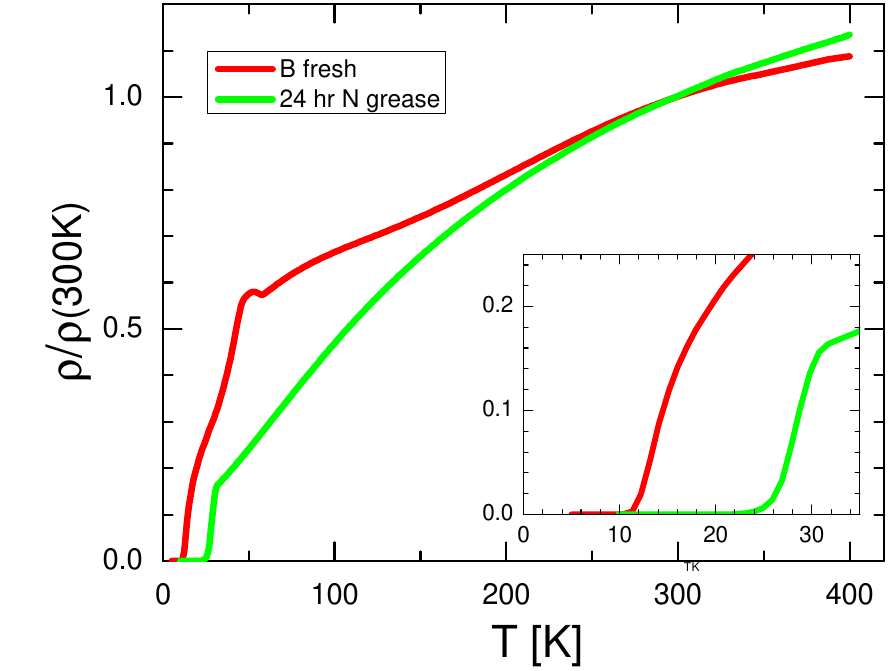}
\caption{(Color online) (Top panel) The evolution of $\rho_a(T)$ of a Na$_{1-\delta}$FeAs crystal during oxidative deintercalation of Na on exposure to Apiezon N-grease. Resistivity of the sample at room temperature, $\rho_a (300K)$, increases slightly, presumably due to microscopic sample degradation and the formation of cracks. The bottom panel shows the same data, plotted with normalized resistivity scale, $\rho / \rho (300K)$. The inset in the bottom panel zooms in on the superconducting transition range, showing increase of the superconducting transition temperature from $T_{c,ons}$=15~K and $T_{c,off}$=12~K in the fresh state to $T_{c,ons}$=30~K and $T_{c,off}$=26~K in the grease exposed state.
}%
\label{resistivitygrease}%
\end{figure}

Figure \ref{resistivitygrease} shows electrical resistivity of samples covered with Apiezon N-grease after initial preparation. Between two measurements sample was left on a PPMS puck in air for 24 hours. Temperature-dependent sample resistivity is shown in the top panel of Fig.~\ref{resistivitygrease}, while the same data on a normalized resistivity scale, $\rho/\rho (300K)$, and a zoom of the superconducting transition are shown in the bottom panel. Exposure of the sample to Apiezon N-grease moderately increases $\rho _a(300K)$ and produces a sample of much better quality judging by a superconducting transition, with $T_{c,ons}=$30~K and $T_{c,off}=$26~with actual zero resistance achieved at 24~K.

\section{Discussion}

\subsection{Structural/magnetic ordering and resistivity}

\begin{figure}[tbh]%
\centering
\includegraphics[width=8cm]{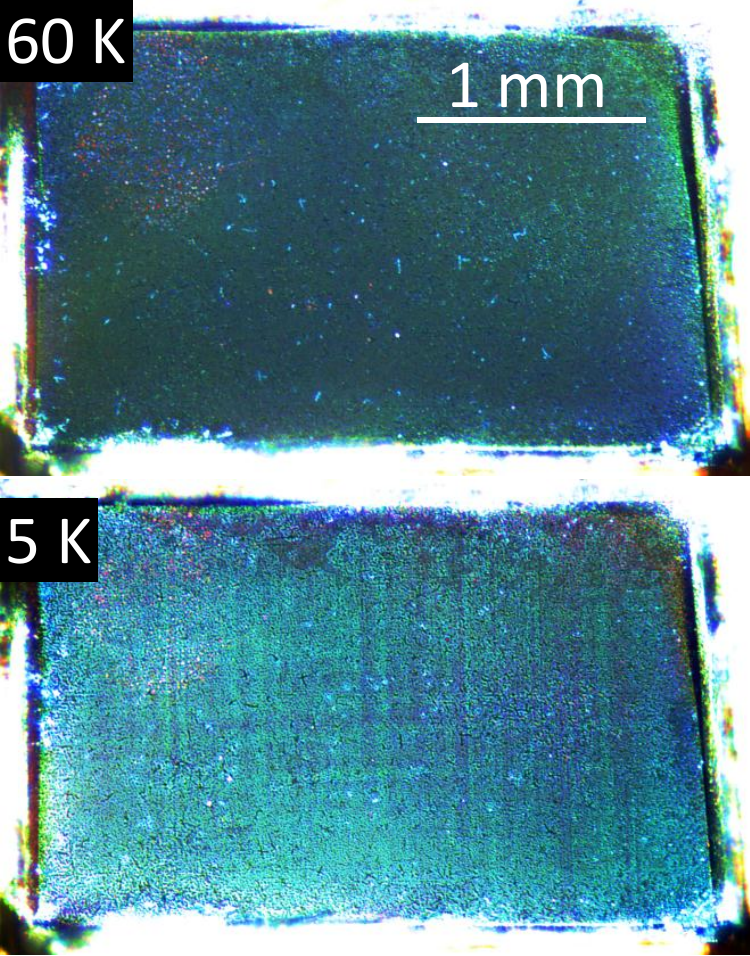}
\caption{Polarized light images of the fresh cleaved surface of a Na$_{1-\delta}$FeAs sample, taken above structural transition at 60~K (top panel) and at 5~K, the base temperature of our experiment (bottom panel). The image shows a pattern of structural domains, with domain walls running along the [100] tetragonal direction. }%
\label{domains}%
\end{figure}

Figure \ref{domains} shows polarized light images of the freshly cleaved surface of the sample used in DC magnetization measurements shown in Fig.~\ref{mag}. The top panel was obtained at 60~K (above the magnetic/structural transition), whereas the bottom panel shows an image taken at 5~K (well below the transitions). It reveals a clear pattern of structural domains, providing direct evidence for the occurrence of the tetragonal to orthorhombic structural transition in our samples. Considering the upturn anomaly in $\rho (T)$ at 56~K (structural transition) and downturn anomaly at 40 K (magnetic transition) measured on another piece of the same crystal, Fig.~\ref{resistivitygrease}, we conclude that freshly cleaved samples exhibit bulk coexistence of superconductivity, magnetism and orthorombic structural distortion.

On the other hand, resistivity in the fresh samples rapidly decreases below a temperature of the magnetic transition, showing that magnetic fluctuations contribute significantly to the inelastic scattering above $T_m$.\cite{detwinning}

\subsection{ Evaluation of the doping level change}

In Na$_{1-\delta}$FeAs, similar to BaFe$_2$As$_2$, doping can be accomplished by substitution of Fe by Co \cite{NaFeAsPRL104}, P-substitution of As \cite{P-NaFeAs}, as well as pressure.\cite{pressure-NaFeAs} Because loss of Na is equivalent to Co substitution, both adding extra electron, it is natural to expect that similar compositional variations are involved. In Fig.~\ref{phaseD} we compare the phase diagram of BaCo122 with that of NaFeAs, using the structural transition temperature as a matching criterion. This way Na$_{1-\delta}$FeAs in its ``fresh'' state corresponds to $x$=0.048 in BaCo122, while at optimal doping Na$_{1-\delta}$FeAs corresponds to $x$=0.07 in BaCo122. This difference, $\Delta x$=0.022 is very close to the actual Co-doping required to induce the highest $T_c$=25~K in Na$_{1-\delta}$FeAs.\cite{NaFeAsPRL104} This comparison suggest that $\delta$ in our samples is of the same order, $\delta \approx$0.02.

\subsection{Evolution of the temperature-dependent resistivity in 111 compounds}

\begin{figure}[tbh]%
\includegraphics[width=8cm]{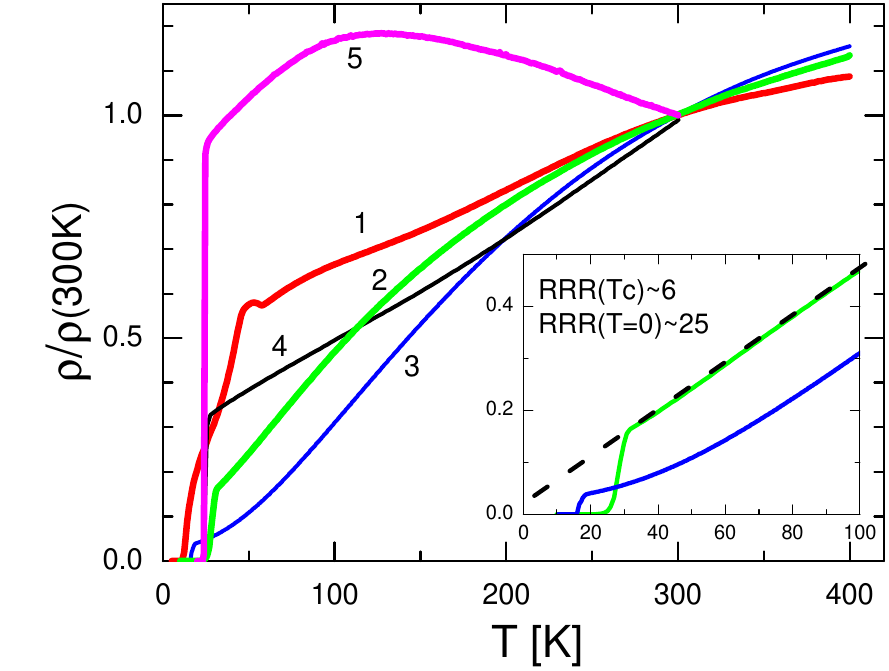}%
\caption{(Color online) The evolution of the temperature-dependent resistivity, plotted on a normalized resistivity scale $\rho/\rho(300K)$, in a 111 system. The data for parent NaFeAs (curve 1), grease treated Na$_{1-\delta}$FeAs (curve 2) and LiFeAs (curve 3) are representative of the underdoped, optimally doped and overdoped regimes, respectively. The inset shows a zoom of the low-temperature range for curves 2 and 3, the line is a linear fit through $\rho(T)$ of grease-deintercalated NaFeAs. For reference we show the in-plane (curve 4) and inter-plane (curve 5) temperature-dependent resistivity of optimally doped Ba(Fe$_{1-x}$Co$_x$)$_2$As$_2$, $x$=0.074, showing $T$-linear resistivity at low temperatures and a pseudogap crossover for $\rho_c(T)$ \cite{pseudogap}.}%
\label{NavsLi}%
\end{figure}

 Figure \ref{NavsLi} shows temperature-dependent resistivity in a 111 system. We plot the data for fresh NaFeAs, grease treated optimally doped Na$_{1-\delta}$FeAs and LiFeAs, representative of the overdoped compositions. For comparison, we show temperature dependence in optimally doped BaCo122. Note that despite macroscopic damage of the crystal during deintercalation, non-stochiometric composition shows quite a high residual resistivity ratio, with $\rho(300K)/\rho(T_c)\approx$6 and $\rho(300K)/\rho(0) \approx $25. This is notably higher than in BaCo122 with $\rho(300K)/\rho(T_c)\approx$3, suggestive that doping into the Fe sites introduces much stronger disorder than the formation of Na vacancies. As can be seen from the inset in Fig.~\ref{NavsLi}, at the optimal doping, $\rho(T)$ in Na111 is purely $T-$linear above $T_c$, which should be contrasted with a quadratic temperature dependence observed in LiFeAs.\cite{HyunsooLiFeAs,TanatarLiFeAs} In the case of Na111 the temperature range of $T$-linear dependence is bound from the high-temperature side by a downward slope change on heating, similar to the case of a hole doped BaK122.

Our findings of doping-independent resistivity at high temperatures, which are particularly clear for air treated samples, are very difficult to explain in a model of additive contributions of two different bands into conductivity. We suggest that a broad crossover in $\rho (T)$ at around 200~K has a similar origin as a broad crossover in $c$-axis transport in BaCo122, where this effect correlates well with the NMR observations of a pseudogap and with the domain of $T-$linear magnetic susceptibility.\cite{pseudogap} It is interesting that the pseudogap crossover does not affect in-plane transport in electron-doped BaCo122, but strongly affects electron-doped Na$_{1-\delta}$FeAs and hole doped BaK122. Finally, we point out a very pronounced branching in the temperature-dependent resistivity, coinciding with the end of the pseudogap resistivity crossover. This feature suggests that the electronic structure of the compounds is affected at temperatures much higher than the temperatures of structural and magnetic transitions, a feature hard to reconcile with simple spin density wave ordering models. It brings some similarity to a pronounced strain-induced anisotropy well above structural and magnetic transitions \cite{Fisher} in the equivalent doping range in BaCo122.

\section{Conclusions}

We used the naturally occurring oxidative deintercalation procedure to fine - tune the doping level of the Na$_{1-\delta}$FeAs system from underdoped to optimally doped. Analyzing DC and AC magnetization, direct optical imaging and temperature - dependent resistivity, we conclude that freshly cleaved samples exhibit bulk coexistence of superconductivity, magnetism and orthorombic structural distortion and can be placed on the underdoped side of the effective $T-x$ phase diagram. The oxidative deintercalation results in a shift of the composition towards the effective optimal doping. Measurements of the in-plane resistivity show that the difference between fresh ($\delta = 0$) and non-stoichiometric states of the crystals starts at about 200~K, much higher than the temperature of the structural/magnetic transitions, $T_s \sim 56$~K and $T_m \sim 40$~K. This temperature is of the same order as the temperature of a broad crossover, found in resistivity measurements in all 111 compounds, which we relate to the formation of a pseudogap. At the optimal doping, the resistivity of Na$_{1-\delta}$FeAs shows an extended range of $T-$linear behavior, in line with the expectations of quantum-critical scenario \cite{Monthoux}, however, this range is bound on the high-temperature side by the onset of a pseudogap crossover.

\section{Acknowledgements}

We thank Seyeon Park for her help with the dipper measurements.
Work at the Ames Laboratory was supported by the Department of Energy-Basic Energy Sciences under Contract No. DE-AC02-07CH11358. The single crystal growth effort at UT is supported by U.S. DOE BES under Grant No. DE-FG02-05ER46202 (P.D.).


\end{document}